%% file: main-arxiv.tex
\documentclass[acmtog]{acmart}

\AtBeginDocument{%
  }

\renewcommand\footnotetextcopyrightpermission[1]{} 

\acmSubmissionID{PAPERS\_1060}

\input{utils/_def}

\usepackage{listings}
\usepackage{xcolor}
\input{utils/math-macro}

\begin{document}

\title{NeuralPVS: Learned Estimation of Potentially Visible Sets}

\author{Xiangyu Wang}
\email{xiangyu.wang@visus.uni-stuttgart.de}
\orcid{0009-0005-0067-2688}
\affiliation{%
  \institution{University of Stuttgart} 
  \city{Stuttgart}
  \country{Germany}
}

\author{Thomas Köhler}
\email{t.koehler@tugraz.at}
\orcid{0009-0004-2685-0502}
\affiliation{%
  \institution{Graz University of Technology}
  \city{Graz}
  \country{Austria}
}

\author{Jun Lin Qiu}
\email{qiu@student.tugraz.at}
\orcid{0009-0002-2782-1549}
\affiliation{%
  \institution{Graz University of Technology}
  \city{Graz}
  \country{Austria}
}

\author{Shohei Mori}
\email{s.mori.jp@ieee.org}
\orcid{0000-0003-0540-7312}
\affiliation{%
  \institution{University of Stuttgart}
  \city{Stuttgart}
  \country{Germany}
}

\author{Markus Steinberger}
\email{markus.steinberger@icg.tugraz.at}
\orcid{0000-0001-5977-8536}
\affiliation{%
  \institution{Graz University of Technology}
  \city{Graz}
  \country{Austria}
}

\author{Dieter Schmalstieg}
\email{dieter.schmalstieg@visus.uni-stuttgart.de}
\orcid{0000-0003-2813-2235}
\affiliation{%
  \institution{University of Stuttgart}
  \city{Stuttgart}
  \country{Germany}
}

\renewcommand{\shortauthors}{Wang et al.}


\begin{abstract}
Real-time visibility determination in expansive or dynamically changing environments has long posed a significant challenge in computer graphics. Existing techniques are computationally expensive and often applied as a precomputation step on a static scene. We present NeuralPVS, the first deep-learning approach for visibility computation that efficiently determines from-region visibility in a large scene, running at approximately 100 Hz processing with less than $1\%$ missing geometry. This approach is possible by using a neural network operating on a froxelized representation of the scene. The network's performance is achieved by combining sparse convolution with a 3D volume-preserving interleaving for data compression. Moreover, we introduce a novel repulsive visibility loss that can effectively guide the network to converge to the correct data distribution. This loss provides enhanced robustness and generalization to unseen scenes. Our results demonstrate that NeuralPVS outperforms existing visibility methods in terms of both accuracy and efficiency.
\end{abstract}



\begin{teaserfigure}
  \includegraphics[width=\textwidth]{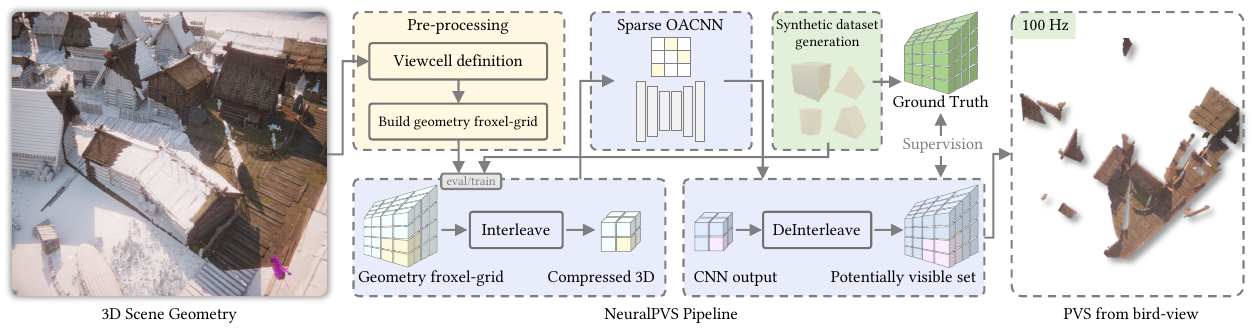}
  \caption{Overview of the NeuralPVS pipeline. The left side illustrates the overall system and task, where the camera is colored purple, and the white rendering indicates geometry invisible to the camera. A froxelized representation of the input scene is fed into the neural network with interleaving layers and outputs the potentially visible set (PVS) in froxelized form, as displayed in the middle. The network is trained with pairs consisting of a froxelized scene and the corresponding ground-truth PVS in froxelized form. The network runs at 100 Hz (10 ms per frame) on the GPU and generates less than $1\%$ error rate, without introducing noticeable artifacts in the rendered images. The right side shows the rendered PVS of the frame from a bird's-eye view.}
  \Description{}
  \label{fig:teaser}
\end{teaserfigure}

\maketitle

\section{Introduction}

Visibility computation is one of the fundamental problems in computer graphics, as it enables a broad spectrum of applications such as shadow mapping, light-field rendering, global illumination, or collision detection. A common approach is to determine a potentially visible set (PVS), which contains an approximation of the visible portion of a scene ~\cite{teller_visibility_1991}. Many methods have been proposed for computing a PVS~\cite{cohen-or_survey_2003}, either from a single viewpoint or from a region (a so-called \textit{viewcell}). Commercial game engines frequently determine a from-point PVS before shading to reduce shading load. Computing a from-region PVS has much broader use for applications such as frame extrapolation, prefetching or streaming rendering. Unfortunately, algorithms for from-region PVS have an inherently high computational complexity~\cite{durand_3d_2002}, and most algorithms rely on precomputation~\cite{bittner_adaptive_2009, mattausch_adaptive_2006}, including solutions used in commercial game engines such as Unreal~\cite{epic_games_precomputed_2022}. Some recent work \cite{hladky_camera_2019, koch_guided_2021, voglreiter_trim_2023, kim_potentially_2023} achieves promising online visibility computation, but the limitations of the underlying algorithms still restrict their applicability, especially for high resolutions and scenes with high geometric complexity.

In recent years, neural networks have revolutionized many areas of computer graphics, such as material representation \cite{sztrajman_neural_2021,dou_real-time_2024}, denoising \cite{bako_kernel-predicting_2017, chaitanya_interactive_2017}, and global illumination \cite{diolatzis_active_2022,ren_lightformer_2024,zheng_neural_2024}. Even complex and ill-posed problems can be successfully handled by neural networks, if enough training data is available for supervised learning. In computer graphics, the problem of procuring enough training data can often be solved by generating synthetic images or datasets. With sufficient optimization, the resulting trained network may be able to run on a GPU at rates that are competitive to conventional algorithms. For example, NVIDIA DLSS~\cite{liu_dlss_2020} uses a neural network for frame upsampling.

In this paper, we explore the use of a convolutional neural network (CNN) to compute a from-region PVS. We rasterize the given scene into a froxel grid, i.e., a frustum-aligned grid expressed in normalized device coordinates~\cite{evans_learning_2015}. Then, we compute a ground-truth PVS as another grid in the same space, by sampling visibility from many viewpoints in the viewcell. Pairs of geometry grid and ground-thruth PVS are derived from synthetic scenes and used for training. At inference time, the network predicts the PVS for the scene as seen from the current viewpoint.

Our approach, called \textit{NeuralPVS}, makes use of the most recent refinements to CNN architectures, which rely on sparse and adaptive convolutions~\cite{peng_oa-cnns_2024}. To boost the speed of our network to real-time performance, we extend the network with 3D interleaving of the froxelized input data. Our end-to-end training pipeline allows the model to learn the visibility patterns directly from synthetic scenes, which consists of basic geometric primitives with varying complexity and occlusion patterns. The trained network can be applied to arbitrary scenes without any need for fine-tuning. This design makes our network a drop-in replacement for a conventional PVS generator. We also introduce a novel repulsive visibility loss function that encourages the model to focus on the most relevant regions and stay away from invisible regions. 
Our main technical contributions can be summarized as follows:
\begin{itemize}
    \item We propose a novel neural network for from-region PVS computation, which has significantly improved accuracy and speed compared to existing methods.
    \item We explore 3D volumetric interleaving and repulsive visibility loss to enhance the efficiency of the neural network, while preserving the accuracy of visibility estimation.
    \item We develop a rendering pipeline and application with our proposed method, which demonstrates the effectiveness of our approach through extensive experiments on large scenes, showing that our method outperforms existing state-of-the-art methods in terms of both accuracy and efficiency.
\end{itemize}

\section{Background}

We begin by surveying prior work on PVS computation, distinguishing between offline methods and those designed for real-time execution. We then review recent advances in integrating deep learning into the traditional graphics pipeline, highlighting how neural architectures have been employed to solve graphics tasks. Finally, we discuss contemporary techniques in 3D geometric learning that inform the backbone network architecture adopted in our work.

\subsection{PVS computation}


For performance reasons, computing PVS offline for the region is a common practice to speed up rendering. The space of possible viewpoints can be subdivided into static viewcells, and the PVS per viewcell can be computed and stored \cite{teller_visibility_1991}. Early offline PVS computation methods are restricted to handling 2.5D geometry~\cite{wonka_visibility_2000} and often imposed additional constraints such as watertight geometry \cite{schaufler_conservative_2000} or particular data distribution schemes \cite{cohen-or_conservative_1998}. Subsequent research aimed to remove these restrictions and address more general geometric scenarios using various techniques based on rasterization~\cite{nirenstein_hardware_2004, bittner_fast_2005} or raytracing~\cite{bittner_adaptive_2009}. The main problem of offline methods remains that they can only work with static scenes.

Modern graphics applications, \eg, 3D streaming and dynamic 3D scenes, call for PVS computation methods that can operate in real time. From-point methods often rely on depth buffering to test and store occlusions. For example, rendering engines can use a geometry pre-pass to establish visible fragments~\cite{burns_visibility_2013}. If a full-resolution geometry pre-pass is considered too expensive, an efficient software rasterizer working at lower resolution~\cite{hasselgren_masked_2016} or a hierarchical depth buffer~\cite{greene_hierarchical_1993} can be used. The latter can be accelerated with hardware occlusion queries, although the need for GPU synchronization limits scalability~\cite{mattausch_adaptive_2006}. Another approach reduces the required geometry processing by replacing complex scene geometry with simpler occlusion proxies~\cite{koltun_virtual_2000}. Finding good proxies is still an active research topic~\cite{tan_differentiable_2025}.

Only a few methods address online computation of a from-region PVS. Simple solutions can try to heuristically sample multiple predicted camera positions~\cite{mueller_shading_2018, hladky_quadstream_2022, reinert_proxyguided_2016}, but this approach is not very scalable.

The camera offset space of Hladky et al.~\shortcite{hladky_camera_2019} uses per-pixel linked lists as a scene representation to determine the PVS. Maintaining sorted lists is expensive and difficult to scale to high resolutions. 
Koch et al.~\shortcite{koch_guided_2021} use hardware ray-tracing to find visible surfaces. Their method is stochastic, and its convergence depends on the speed of the raytracing hardware. Kim and Lee~\shortcite{kim_potentially_2023} propose track disocclusions through depth peeling. Scalability of depth peeling is limited by the need to repeatedly rasterize the scene. Voglreiter et al.~\shortcite{voglreiter_trim_2023} combine traversal of a coarse octree in world space with k-buffering in image space. Peeling octree layers instead of traditional depth peeling allows them to build the PVS with a single geometry pass. However, the scalability of their method is still affected by the need for GPU synchronization after each layer.

Recently, a disocclusion-based approach was proposed \cite{kunzel_potentially_2025}, introducing the disocclusion buffer as a sparse, layered representation that allows order-independent and fully parallel computation of PVS, achieving speed-ups with comparable accuracy. Although both this concurrent work and ours advance PVS generation, our neural network-based approach will benefit even more from future GPU advancements, as improved hardware directly accelerates inference performance.

Our method builds on the observation that visibility can be efficiently expressed in a froxelized view frustum. A froxel grid requires only one bit per froxel to indicate a disoccluded region. Such a froxel grid can be built much more efficiently than linked lists, k-buffers or depth peeling layers. We use the froxel grid as input to a CNN, which has a run-time cost that is only dependent on the froxel grid resolution. This resolution is independent of both the target image resolution and the complexity of the scene.

\subsection{Deep learning in graphics}

With the ability to learn and model complex mappings, deep learning is increasingly gaining recognition as a new approach to overcome bottlenecks in computer graphics because of its predictable computational cost and speed.

Deep Shading \cite{nalbach_deep_2017} is one of the earliest works to introduce a CNN into graphics tasks to compute the shading effect on pixels from shading buffers. A CNN has also been used to efficiently denoise Monte Carlo rendering \cite{bako_kernel-predicting_2017, chaitanya_interactive_2017}. Similar neural network-based techniques are widely used in post-processing, such as by DLSS \cite{liu_dlss_2020}. BRDF methods based on deep learning \cite{hu_deepbrdf_2020, sztrajman_neural_2021, dou_real-time_2024} use neural layers to represent BRDF to replace the large tabulated dataset and can achieve a better trade-off between memory and quality. Neural global illumination \cite{diolatzis_active_2022, ren_lightformer_2024,zheng_neural_2024} can now compete in quality for dynamic lighting simulation \cite{ren_lightformer_2024} and deliver convincing results on dynamic geometry \cite{zheng_neural_2024}.

To the best of our knowledge, no existing work addressed visibility computing with a neural network. In this paper, we will focus on applying deep learning methods to the field of PVS computation.

\subsection{3D geometric learning}

Our PVS estimation network combines several key techniques from 3D geometric learning, including sparse 3D convolution and the latest improvements to the convolutional network design.

Classical 3D convolutional networks \cite{dai_scannet_2017, tchapmi_segcloud_2017} use dense representation to learn 3D geometry patterns. Dense representations have extremely high memory consumption and lead to slow inference speed. Most geometric models have ample empty space and do not require convolutions to be applied in the empty areas. Therefore, a sparse CNN architecture is usually preferred for this kind of problem.

Instead of storing the entire space as one tensor, a sparse 3D convolution network stores its data, the so-called sparse tensor, either in an octree \cite{riegler_octnet_2017} or in a hash table \cite{graham_submanifold_2017,choy_4d_2019, spconv_contributors_spconv_2022, chen_focal_2022}. Sparse 3D convolutions on hash tables are optimized by placing the kernel center at the activated positions \cite{graham_submanifold_2017}, dilating to the activated position's neighbors \cite{choy_4d_2019}, dynamically dilating the reception field \cite{chen_focal_2022}, or combinations of these factors \cite{choy_4d_2019,spconv_contributors_spconv_2022}. Recent work~\cite{peng_oa-cnns_2024} on an omni-adaptive convolutional neural network (OA-CNN) demonstrates that advances in transformer networks can be retrofitted to CNN architectures by adding adaptive receptor fields and a form of self-awareness. Our method benefits from the speed and efficiency afforded by these architectures, since a volumetric scene typically has an occupancy of less than 5\% of the froxels.

 \begin{figure}[htbp]
    \centering
    \begin{subfigure}[t]{0.23\textwidth}
        \centering
        \includegraphics[height=2.7cm,keepaspectratio]{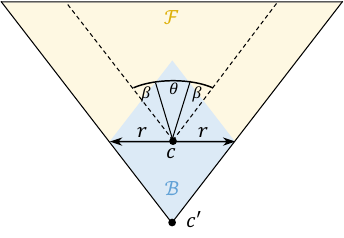}
        \caption{Viewcell}
        \label{fig:viewcell}
    \end{subfigure}
    \begin{subfigure}[t]{0.23\textwidth}
        \centering
        \includegraphics[height=2.7cm,keepaspectratio]{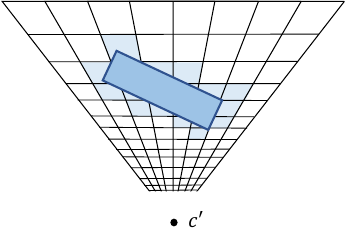}
        \caption{Geometry froxel-grid}
        \label{fig:gv}
    \end{subfigure}  
  \caption{(a) A frustum enclosing all primitives that are potentially visible from a viewcell (blue area) with radius $r$ around the current viewpoint $\mathbf{c}$ is created by displacing the viewpoint backwards to $\mathbf{c}'$. (b) The scene primitives are conservatively rasterized into a regular grid.}
  \label{fig:viewcell_gv}
\end{figure}

\section{Method}
\label{sec:method}

\begin{figure*}
    \centering
    \includegraphics[width=\linewidth]{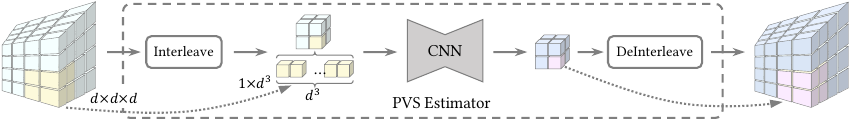}
    \caption{NeuralPVS pipeline. For each viewcell, the scene’s geometry is froxelized into a GV, which is input to the PVS estimator network. A 3D interleaving function first compresses the GV channels; a CNN then predicts the visible part of the geometry grid; afterwards, a 3D deinterleaving function reconstructs the full \pvv. Geometric primitives in froxels marked invisible in the \pvv are culled from all further rendering computations.}
    \label{fig:pipeline}
\end{figure*}

Our goal is to significantly improve the performance of PVS computation by replacing previous algorithms operating on analytic or sampled geometry with a robust neural network operating on a froxelized scene representation. The network is pre-trained with synthetic geometry grids corresponding to random scenes that loosely resemble the structure of the target scenes. A key advantage of using a CNN comes from the fact that converting the polygonal scene into a froxel grid of fixed resolution makes the time needed to compute a PVS largely independent of geometric scene complexity. Overall, our pipeline proceeds as follows (\figurename~\ref{fig:pipeline}):
\begin{enumerate}
  \item \textbf{Viewcell definition}: build a geometry grid $G$ by rasterizing the scene into $G$
  \item \textbf{\pvv generation}: a single forward pass through the neural network infers the PVS
  \item \textbf{Novel view synthesis}: rasterize only primitives contained in the PVS
\end{enumerate}
Step 3 is repeated until the camera leaves the current viewcell, then the process is restarted with step 1.

\subsection{Preliminaries}
\label{sec:method-overview}

We represent the scene geometry by a grid $G(\mathbf{x}), \mathbf{x}\in\mathbb{X}$, in normalized device coordinates, which is 1 if the froxel at $\mathbf{x}$ is occupied, and 0, otherwise. $G$ is defined over a discrete domain $\mathbb{X} \equiv [1..N_x]\times[1..N_y]\times[1..N_z]$, which splits the view frustum into a grid of discrete froxels.

The potentially visible set can be defined as the union of visible polygons for all viewpoints in a cell \cite{airey_towards_1990}. Based on the definition, we represent the PVS in volumetric form as a grid $V(\mathbf{x})$, which is $1$ for visible froxels, and $0$ otherwise. Let $N=N_x\cdot N_y\cdot N_z$. Our goal is to learn a mapping from $G$ to the ground-truth per-froxel visibilities
    $\hat V: \{0,1\}^{N} \;\rightarrow\; \{0,1\}^{N}.$
%
We train a 3D convolutional network $\mref{neu_func}$ that produces per-froxel probabilities and obtain ${V} = \mathbf{1}(\mref{neu_func}(G))$, where $\mathbf{1}$ is an indicator function $[{V}(\mathbf{x}) \ge \mref{pvv_thres}]$ with a threshold $\mref{pvv_thres}$.

\subsection{Pre-processing}
\label{sec:method-preprocessing}
\label{sec:method-viewcell}

Our definition of the viewcell $\mathcal{B}$ (Figure~\ref{fig:viewcell_gv}a) considers a lateral motion of the camera from an original viewpoint $\mathbf{c}$ up to radius $r$~\cite{wonka_instant_2001}. If we assume that the camera has a field of view of $\theta$, the view frustum $\mathcal{F}$ associated with a viewpoint $\mathbf{c}'$ displaced backward by $r/\tan(\theta/2)$ encloses the PVS associated with $\mathcal{B}$. To accommodate camera rotations up to a maximum angle of $\beta$ on either side, we enlarge the field of view to $\theta+2\beta$.

\label{sec:method-gv_gen}

We rasterize the primitives contained in $\mathcal{F}$ into a geometry grid $G$ stored as a 3D texture, according to the concept of froxelization \cite{evans_learning_2015}. Each fragment $(x,y,z)^\top$ in normalized device coordinates is quantized to the froxel coordinates at $\scalebox{0.8}[1]{$\bigl(\lfloor u\,N_x\rfloor,\;\lfloor v\,N_y\rfloor,\;\lfloor w\,N_z\rfloor\bigr)^\top$}$, and the texture at the corresponding location is set to 1 to indicate an occupied froxel. To ensure a gap-free rasterization, we supersample the scene at a resolution of \((sN_x,sN_y,sN_z)\). For storage efficiency, we pack eight consecutive froxels along the \(x\)‐axis into an 8-bit integer using atomic bitwise-OR for concurrent texture writes. For the ground-truth generation during training, the fragments are generated using orthographic projections and then reprojected on the fly into $\mathcal{F}$ to ensure a uniform sample distribution across all distances from the camera. At test time, we prioritize speed and use a conventional perspective projection to generate the samples for the geometry grid.

By construction, \(\mathrm{supp}(\mref{pvv})\subseteq\mathrm{supp}(V)\), which allows immediate occlusion culling: Any fragment mapping to a froxel where \(\hat{V}=0\) can be discarded. As proposed by Hladky et al.~\shortcite{hladky_camera_2019}, the ground truth \pvv is computed by dense sampling of the viewcell. A large number ($M=1000)$ of viewpoints $\mathbf{c}_m\in\mathcal{B}$ is selected. For each viewpoint, the primitives in $\mathcal{F}$ are rendered and a depth buffer is produced. The fragments indicated in the depth buffer are reprojected to the coordinate system of $\mathbf{c}$, and the corresponding froxels in $G$ are marked as occupied. 

\subsection{Neural PVS estimation}

Dense volumetric CNN architectures, such as VNet \cite{milletari_v-net_2016}, are designed for offline operation, such as segmentation of medical scans. Even their sparse variants are too slow for applications in real-time graphics. Therefore, we adopted OA-CNN~\cite{peng_oa-cnns_2024} as a backbone. OA-CNN is built on a highly optimized kernel for sparse convolution. Furthermore, it introduces adaptive receptive fields and dynamically adjusts convolutional kernel weights to deliver performance that reflects modern transformer networks. The network accepts a geometry grid $G$ as input and provides volumetric probabilities $V$ as output. The complete pipeline is described in \Cref{fig:pipeline}.

\paragraph*{3D volume-preserving interleaving}
\label{sec:method-interleaving}

The CNN inference time complexity is linearly dependent on the resolution of the input features, while the number of features has less impact on the speed. To further improve the inference speed, we adopt a 3D volume-preserving interleaving, generalizing the mechanism proposed by Xiao et al.~\shortcite{xiao_deepfocus_2018}. As shown in \Cref{fig:pipeline}, we place an interleaving function $g_d$ before the convolutional layers and a corresponding de-interleaving function after the convolutional layers:
\begin{align} \nonumber
    V &= {g_d^{-1}}(\mref{neu_func}(g_d(G))).
\end{align}
The interleaving function 
\begin{align} \nonumber
    g_d &: \mathbb{R}^{N_x\times N_y\times N_z} \rightarrow \mathbb{R}^{\frac{N_x}{d}\times\frac{N_y}{d}\times\frac{N_z}{d}}
\end{align}
takes as input a grid of $N_x\times N_y\times N_z$ froxels. It divides the grid into blocks of dimension $d\times d\times d$ and stacks the froxels in a block into a one-dimensional feature vector. The de-interleaving function 
\begin{align} \nonumber
    g_d^{-1} &: \mathbb{R}^{\frac{N_x}{d}\times\frac{N_y}{d}\times\frac{N_z}{d}} \rightarrow \mathbb{R}^{N_x\times N_y\times N_z} 
\end{align}
inverts this process. We choose $d\in\{8, 16, 32\}$ for optimal memory alignment. A value $d<8$ is not practical, as the setup overhead becomes too high~\cite{xiao_deepfocus_2018}. The interleaving preserves the relative positional information of the geometry, while shrinking the dimension of the input by a factor of $d^3$. For the typical setup of $d=16$, the first convolution step after interleaving further shrinks the size of each input vector from $d^3$ to around 200 elements, leading to a significantly reduced processing time.

\begin{figure*}[t]
    \centering
    \includegraphics[width=\linewidth]{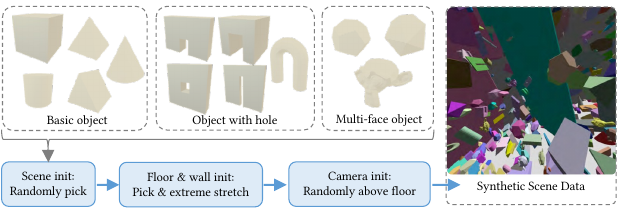}
    \caption{The pipeline of our proposed synthetic 3D data generation. To ensure the data distribution diversity, we applied a multi-step scene generation strategy in order to cover as many occlusion patterns as possible.}
    \label{fig:dataset-generation}
\end{figure*}

\paragraph*{Weighted Dice loss}

We adopt a weighted Dice loss, which weighs false negatives (FN) much more strongly than false positives (FP) by a factor $\alpha$. In our context, false negatives are froxels included in the ground truth PVS, but missing from the predicted PVS. False positives are froxels not included in the ground truth, but included in the predicted PVS. True positives (TP) are froxels which are included both in the ground truth and the predicted PVS, and ground truth positives (GTP) are all froxels marked in the ground truth $\hat V$:
\begin{align} \nonumber
    \mref{neu_tp}(V,\hat V) = \sum_{\mathbf{x}\in\mathbb{X}}{\hat{V}(\mathbf{x})\cdot V(\mathbf{x})}, \enspace \nonumber
    \mref{neu_fp}(V,\hat V) = \sum_{\mathbf{x}\in\mathbb{X}}{\hat{V}(\mathbf{x})\cdot (1-V(\mathbf{x})}),\\ 
    \mref{neu_fn}(V,\hat V) = \sum_{\mathbf{x}\in\mathbb{X}}(1-{\hat{V}(\mathbf{x}))\cdot V(\mathbf{x})}, \enspace \nonumber
    \text{GTP}(\hat V) = \sum_{\mathbf{x}\in\mathbb{X}} \hat V(\mathbf{x}). \nonumber
\end{align}
Following the definition of Taha and Hanbury~\shortcite{taha_metrics_2015}, our modified Dice loss is given as 
\begin{align} \nonumber
    \mref{loss_dice}(V,\hat V) = 1 - \frac{2~\mref{neu_tp}(V,\hat V)}{2~\mref{neu_tp}(V,\hat V) + \alpha\mref{neu_fp}(V,\hat V) + (1-\alpha)\mref{neu_fn}(V,\hat V)}.  \nonumber 
    \label{eq:dice}
\end{align}

\paragraph*{Repulsive visibility loss}
\label{sec:method-rvloss}

Our training data is necessarily very imbalanced, because the number of visible froxels in real scenes is usually between 0.5\% and 10\% and can vary significantly between scenes and viewpoints. This imbalance makes it difficult to achieve convergence, as the network can easily get stuck in a local minimum during training. The conditioning to avoid high false negative counts can lead to overprediction of visibility, or the network never leaves the initial state.

To overcome these imbalance issues, we propose a repulsive visibility loss (RVL) inspired by the work of Wang et al.~\shortcite{wang_repulsion_2018}. Unlike Dice loss, which globally supervises all froxels, RVL aggressively encourages only prediction in local grids where $V=1$ to match the overall distribution of FN/FP.
This loss is given as
\begin{align} \nonumber
    \mathcal{L}_{rv} = \mathcal{L}_{attr}+\mathcal{L}_{rep}, \quad 
    \mathcal{L}_{attr} = 1- \frac{\text{FN}}{\text{GTP}}, \quad \mathcal{L}_{rep}=\frac{\text{FP}}{\text{GTP}},
\end{align}
where $\mathcal{L}_{attr}$ attracts the prediction to match the ground truth, and ${\mathcal{L}}_{rep}$ pushes the prediction away from non-visible areas. We train the model by minimizing a combination of $\mref{loss_dice}$ and $\mathcal{L}_{rv}$:
\begin{align} \nonumber
    \mref{loss} &= \mref{lossw}\cdot\mref{loss_dice}(V, \hat V) + (1 - \mref{lossw})\cdot\mref{loss}_{rv}(V,\hat V).
\end{align}

\section{Implementation}
\label{sec:impl}

\subsection{Datasets}

The training dataset must include a wide range of geometric variations to ensure generalization. To achieve this, we use synthetic data. We randomly place selected objects at varying frequencies and apply random scales and rotations with a consistent distribution across the local object dimensions (\Cref{fig:dataset-generation}).
Each training set is composed of 1000 frames.

Our objects consist of primitive and complex shapes, including cubes, cones, pyramids, cylinders, dodecahedrons, icosahedrons, and round arches, as well as the monkey head model from Blender, a wall with a door-shaped cutout, and a cube with a square hole resembling a window. To simulate architectural structures such as walls, floors, and ceilings, we scale certain objects significantly along two of their three dimensions. Finally, we introduce planes that serve as a global base for the floor, wall, and ceiling, extending across the scene. These planes are initialized to ensure that they remain visible within the viewcell. The viewcell is initialized at the center of the scene, positioned above the floor at a random height, and randomly rotated for variations.

We generate evaluation datasets using widely used scenes, such as Viking Village and Robot Lab, for comparison with previous work. For both the training and evaluation sets, we create pairs of a geometry grid and ground-truth PVS (\Cref{sec:method-gv_gen}).

\subsection{Training and evaluation}

Our neural network implementation is based on OA-CNN \cite{pointcept2023} (see supplementary material for details). For performance reasons, we use an aspect ratio of 1:1 for our viewport. Specifically, we test grid resolutions of $A \times A \times A$ with $A = 256$ and a viewcell size $r \in \{30\ \text{cm}, 60\ \text{cm}, 90\ \text{cm}\}$.

All training was performed on an Oracle Linux 7.9 server with 2 AMD EPYC 7662, 1 TB RAM, and 4 NVIDIA A100 SXM4 GPU; only one GPU was used for training. All evaluations were run on a desktop computer running Oracle Linux 9.5, equipped with AMD Ryzen 9 7900X, 64 GB RAM, and an NVIDIA RTX 5090 GPU. We optimized the model using $\mref{lossw}=0.99$, with an initial learning rate $0.001$, learning rate decay rate $10^{-10}$ and batch size $3$. We performed 200 epochs of training, which took approximately 15 hours for our setup. We used $\tau=0.5$ throughout all experiments.

\subsection{Rendering}

We use Unity's Universal Rendering Pipeline (URP) version 17.0 for all rendering tasks. The main camera is configured with a resolution

of 1024$\times$1024 pixels, a field of view (FOV) of 60°, a near plane of 0.3 m, and a far plane of 1000 m. For the rendering of GV and \pvv, we employ a wider FOV of 90° to account for up to 15° of camera rotation within a viewcell. To avoid visibility gaps, we render these buffers at a higher resolution of 2048$\times$2048.

The predicted view frustum spans up to a threshold (empirically chosen at 30 m). For distances larger than the threshold, we render the far-field scene geometry into a visibility buffer once when the PVS is created, and add the geometry indicated in the ID attachment of the visibility buffer to the PVS directly. This optimization allows


\begin{figure}[H]
    \centering
    \begin{subfigure}[b]{\columnwidth}
        \centering
        \includegraphics[width=\linewidth]{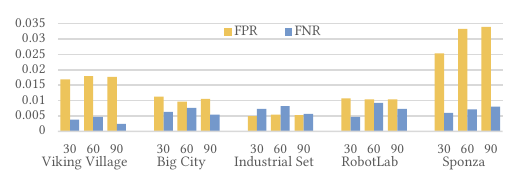}
        \vspace{-2em}
        \caption{Performance metrics for various viewcell size on different scenes. 
        The horizontal axis is grouped by viewcell size $r \in \{30\,\mathrm{cm}, 60\,\mathrm{cm}, 90\,\mathrm{cm}\}$ for each scene.
        Given that our network uses a fixed geometry grid size, performance remains relatively insensitive to changes in $r$.}
        \label{fig:performance-by-r}
    \end{subfigure}
    \begin{subfigure}[b]{\columnwidth}
        \centering
        \includegraphics[width=\linewidth]{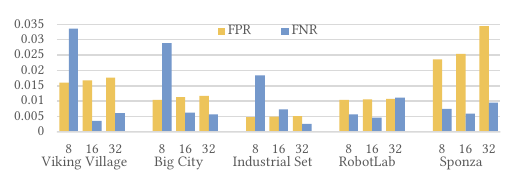}
        \vspace{-2em}
        \caption{Froxel space performance metrics for various interleaving grid size $d$ on different scenes.
        The horizontal axis is grouped by $d \in \{8, 16, 32\}$ for each scene. 
        The performance is affected by changes in $d$; setting $d=16$ gives the overall best performance.}
        \label{fig:performance-by-size}
    \end{subfigure}
    \vspace{-2em}
    \caption{Performance for different viewcell sizes and interleaving (\Cref{sec:method-interleaving}) grid sizes. 
    All metrics were averaged over the entire frame sequence.}
    \label{fig:performance-details-a}
\end{figure}

\vspace{-1.9em}

\begin{figure}[H]
    \centering
    \begin{subfigure}[b]{\columnwidth}
        \centering
        \includegraphics[width=\linewidth]{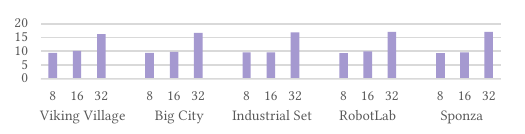}
        \vspace{-2em}
        \caption{Average inference time (ms).}
        \label{fig:result-speed}
    \end{subfigure}
    \begin{subfigure}[b]{\columnwidth}
        \centering
        \includegraphics[width=\linewidth]{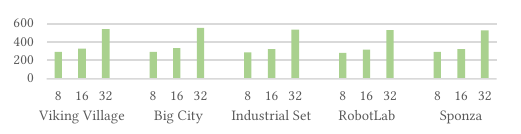}
        \vspace{-2em}
        \caption{Average peak allocated memory (MB).}
        \label{fig:result-memory}
    \end{subfigure}
    \vspace{-2em}
    \caption{Runtime and memory with different interleaving factors $d=\{8,16,32\}$.}
    \label{fig:performance-details-b}
\end{figure}

\vspace{-1.9em}

\begin{figure}[H]
    \centering
    \includegraphics[width=\linewidth]{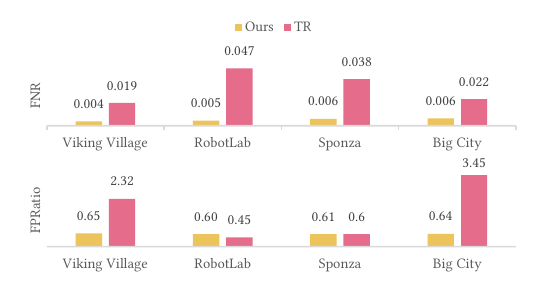}
    \vspace{-2em}
    \caption{Comparison of FNR and FPR (false positive as a multiple of the ground truth) 
    between our results and baseline TR \cite{voglreiter_trim_2023}. 
    In general, our method produces better FNR, similar or better FPR.}
    \label{fig:comparison}
\end{figure}

\noindent the CNN to focus its predictive power in the near- and mid-ranges, where disocclusions are expected.
The ground truth \pvv is sampled using 1000 evenly distributed camera positions per viewcell.

\begin{figure*}[t]
\begin{minipage}[t]{0.59\textwidth}
\centering
\captionof{table}{Comparison of time (ms), memory (MB) and pixel error rate (PER) between our method and Trim Regions \cite{voglreiter_trim_2023}. "Ours" uses $r=30,\ d=16$ and "Ours*" uses $r=30, d=8$ to reduced memory. Pixel error comparison between our method and baseline TR \cite{voglreiter_trim_2023}. Although the baseline has better PER, SSIM comparisons on the image confirm that the errors are minor and almost unnoticeable by humans.}
\label{tab:speed-comparison}
\begin{tabular}{@{}lrr|rrr|rrr@{}}
\toprule
\multirow{2}{*}{Scene} & \multicolumn{2}{c}{ms$\downarrow$} & \multicolumn{3}{c}{MB$\downarrow$} & \multicolumn{2}{c}{PER/\%$\downarrow$} & SSIM$\uparrow$ \\
\cmidrule(lr){2-3} \cmidrule(lr){4-6} \cmidrule(lr){7-8} \cmidrule(lr){9-9}
 & Ours & TR & Ours & Ours* & TR & TR & Ours & Ours \\
\midrule
Viking     & 9.9  & 19.3 & 328.8 & 293.7 & 359.2 & 0.006 & 0.032 & 0.9996 \\
Robot Lab  & 9.9  & 17.2 & 318.0 & 285.1 & 483.8 & 0.030 & 0.255 & 0.9984 \\
Sponza     & 10.1 & 17.8 & 323.4 & 292.0 & 563.9 & 0.021 & 0.151 & 0.9981 \\
City       & 10.3 & 16.4 & 333.5 & 295.3 & 330.5 & 0.006 & 0.142 & 0.9988 \\
\bottomrule
\end{tabular}
\end{minipage}
\hfill
\begin{minipage}[t]{0.39\textwidth}
\centering
\captionof{table}{Ablation study on Viking Village. We use $r=30,\ d=16$. For "No OA-CNN" and "VNet", VNet \cite{milletari_v-net_2016} is used instead of OA-CNN. Time and memory for "No RVLoss" is omitted because the loss does not affect the performance of inference.}
\label{tab:ablation}
\begin{tabular}{@{}lrrrr@{}}
    \toprule & FNR $\downarrow$     & FPR $\downarrow$     & ms $\downarrow$  & MB $\downarrow$ \\
    \midrule
Ours             & 0.00368 & 0.01683 & 10.1  & 328.8  \\
No Interleave & 0.00227 & 0.01940 & 25.4  & 1076.2 \\
No OA-CNN        & 0.01870 & 0.01677 & 19.0  & 509.8  \\
No RVLoss       & 0.00000 & 0.61785 & -     & -      \\
VNet             & 0.03235 & 0.01771 & 326.4 & 2135.9 \\
\bottomrule
\end{tabular}
\end{minipage}
\end{figure*}





\begin{table}[]
\caption{Image space performance metrics SSIM for various interleaving grid size $d$ on different scenes.}
  \label{tab:performance-by-d-ssim}
\begin{tabular}{@{}lllll@{}}
\toprule
\multicolumn{1}{c}{\multirow{2}{*}{Scenes}} & \multicolumn{3}{c}{SSIM $\uparrow$} &  \\ \cmidrule(l){2-5} 
\multicolumn{1}{c}{}                        & d=8    & d=16   & d=32   &  \\ \cmidrule(r){1-4}
Viking Village                              & 0.9990 & 0.9996 & 0.9985 &  \\
City                                        & 0.9991 & 0.9988 & 0.9972 &  \\
Industrial                                  & 0.9963 & 0.9963 & 0.9996 &  \\
Robot Lab                                   & 0.9985 & 0.9984 & 0.9995 &  \\
Sponza                                      & 0.9998 & 0.9981 & 0.9913 &  \\ \bottomrule
\end{tabular}
\end{table}

\section{Results}

\begin{figure*}[t]
  \centering
  \begin{subfigure}{0.19\textwidth}
    \includegraphics[width=\linewidth]{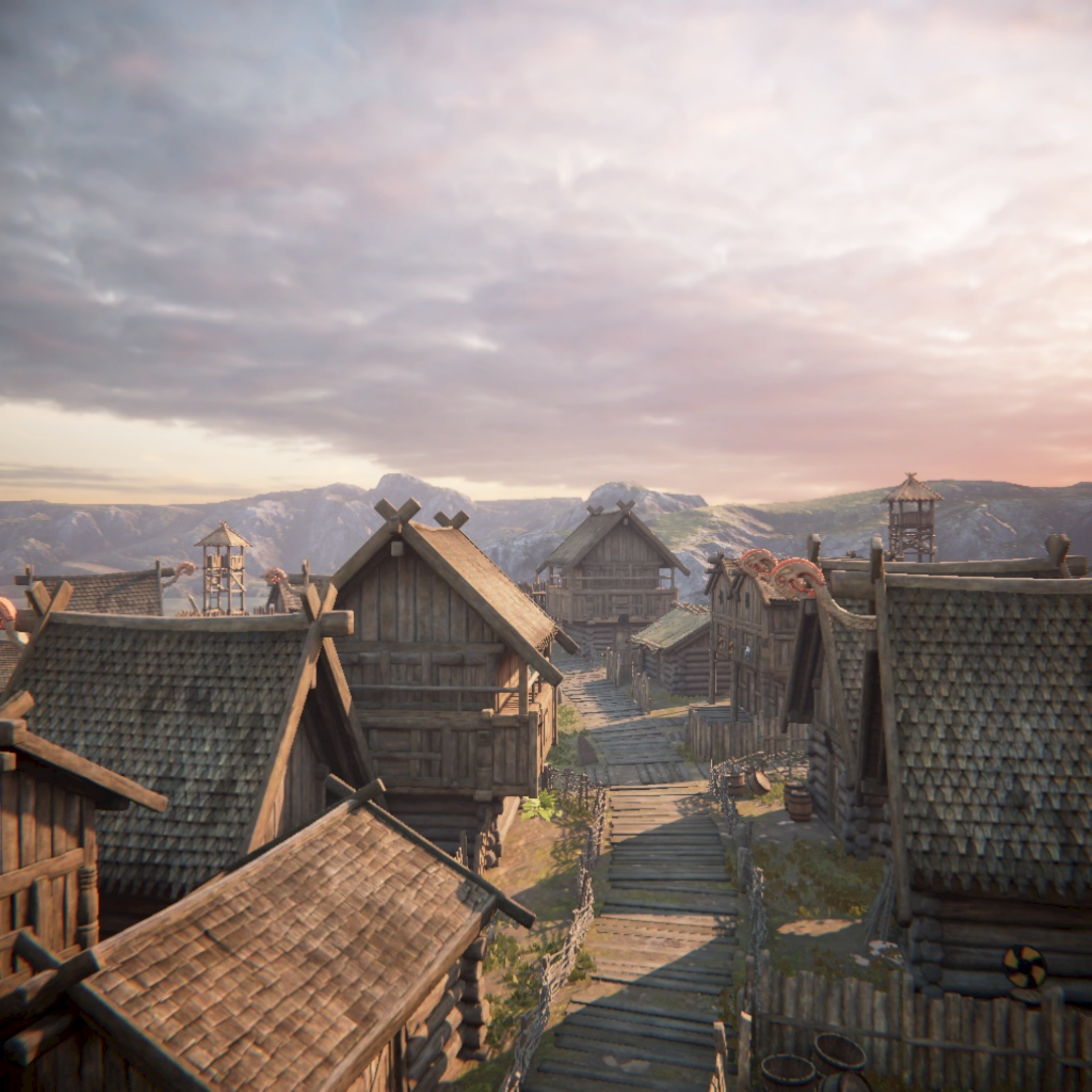}
    \caption*{Viking Village (7.2M)}
  \end{subfigure}
  \hfill
  \begin{subfigure}{0.19\textwidth}
    \includegraphics[width=\linewidth]{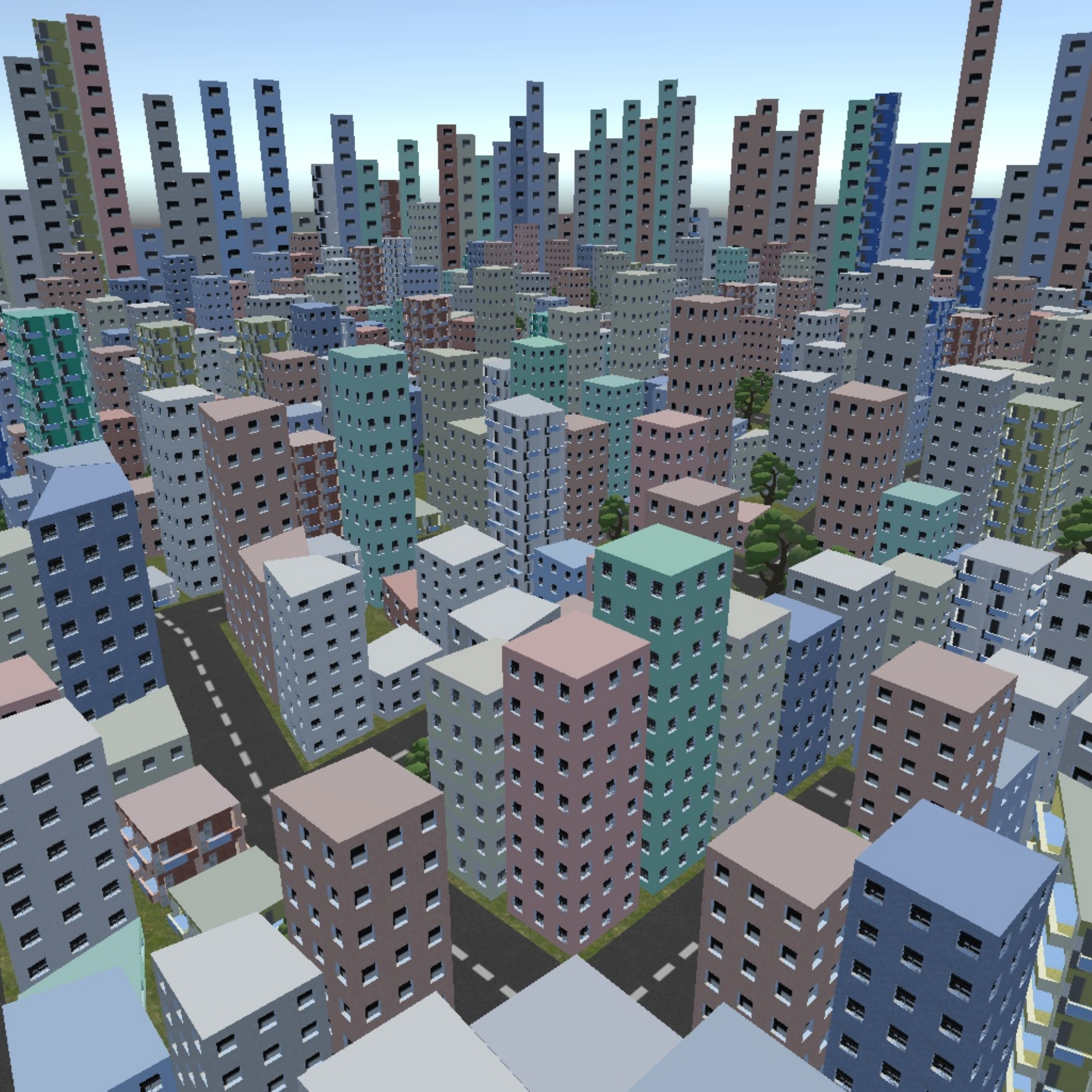}
    \caption*{Big City (15.7M)}
  \end{subfigure}
  \hfill
  \begin{subfigure}{0.19\textwidth}
    \includegraphics[width=\linewidth]{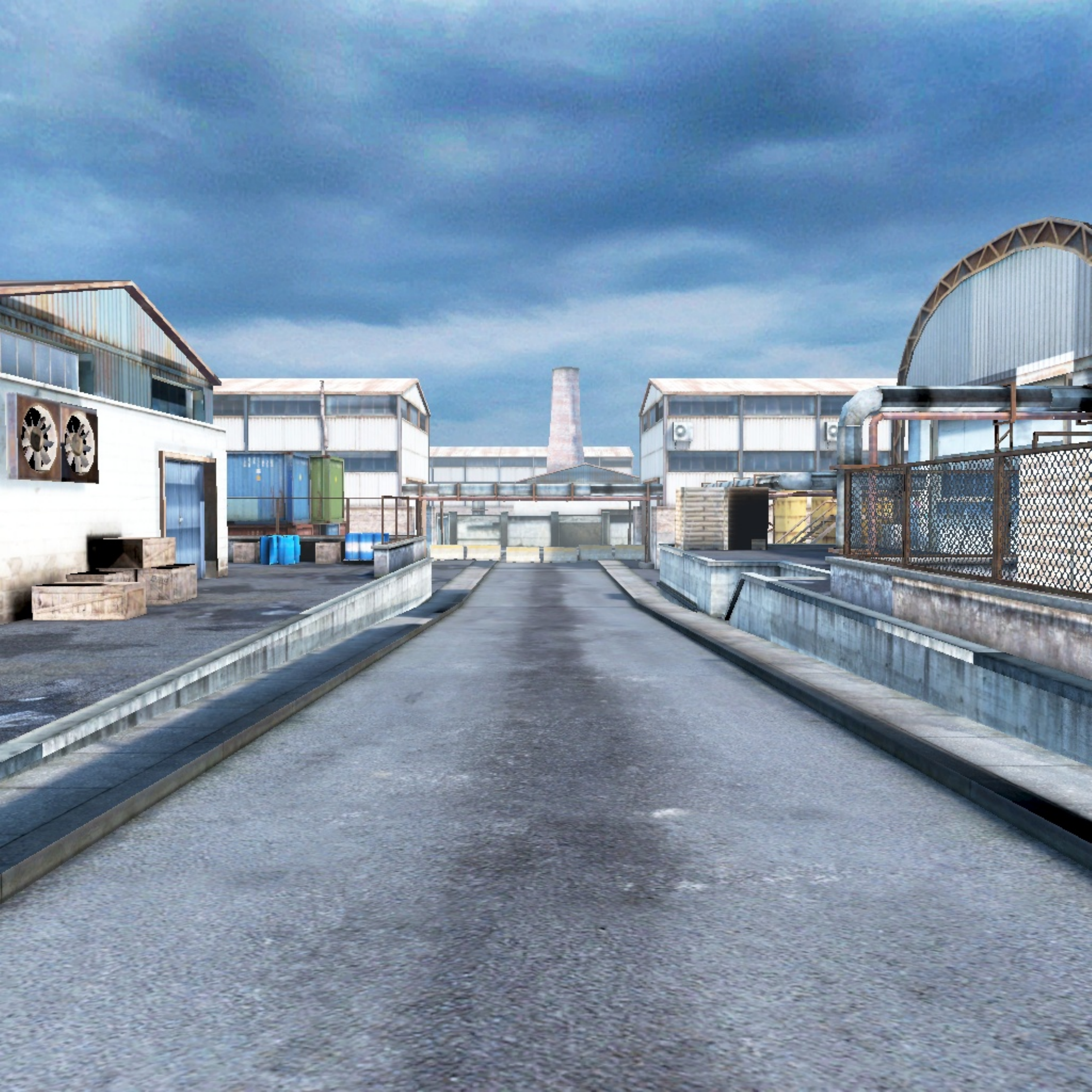}
    \caption*{Industrial Set v3.0 (0.1M)}
  \end{subfigure}
  \hfill
  \begin{subfigure}{0.19\textwidth}
    \includegraphics[width=\linewidth]{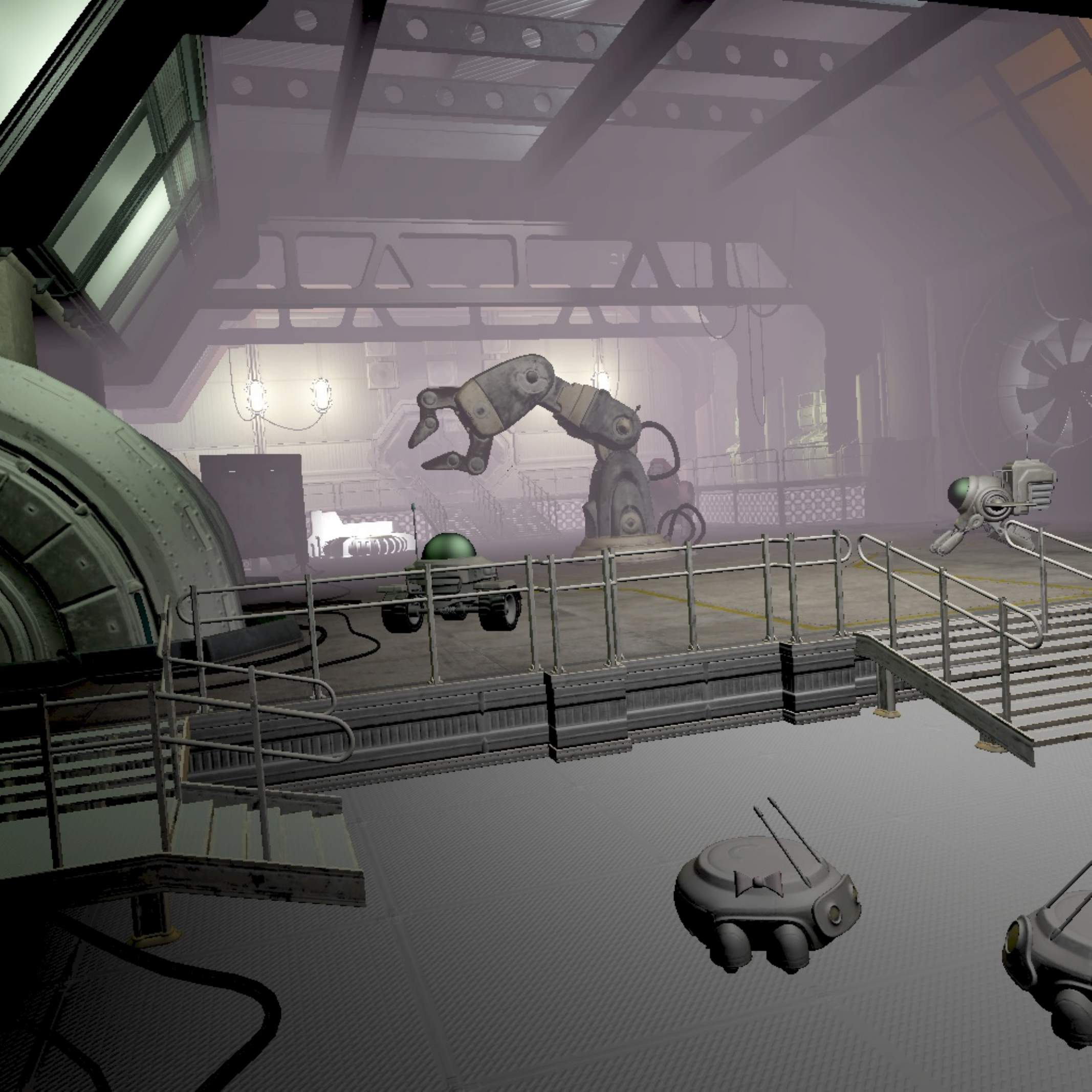}
    \caption*{Robot Lab (0.5M)}
  \end{subfigure}
  \hfill
  \begin{subfigure}{0.19\textwidth}
    \includegraphics[width=\linewidth]{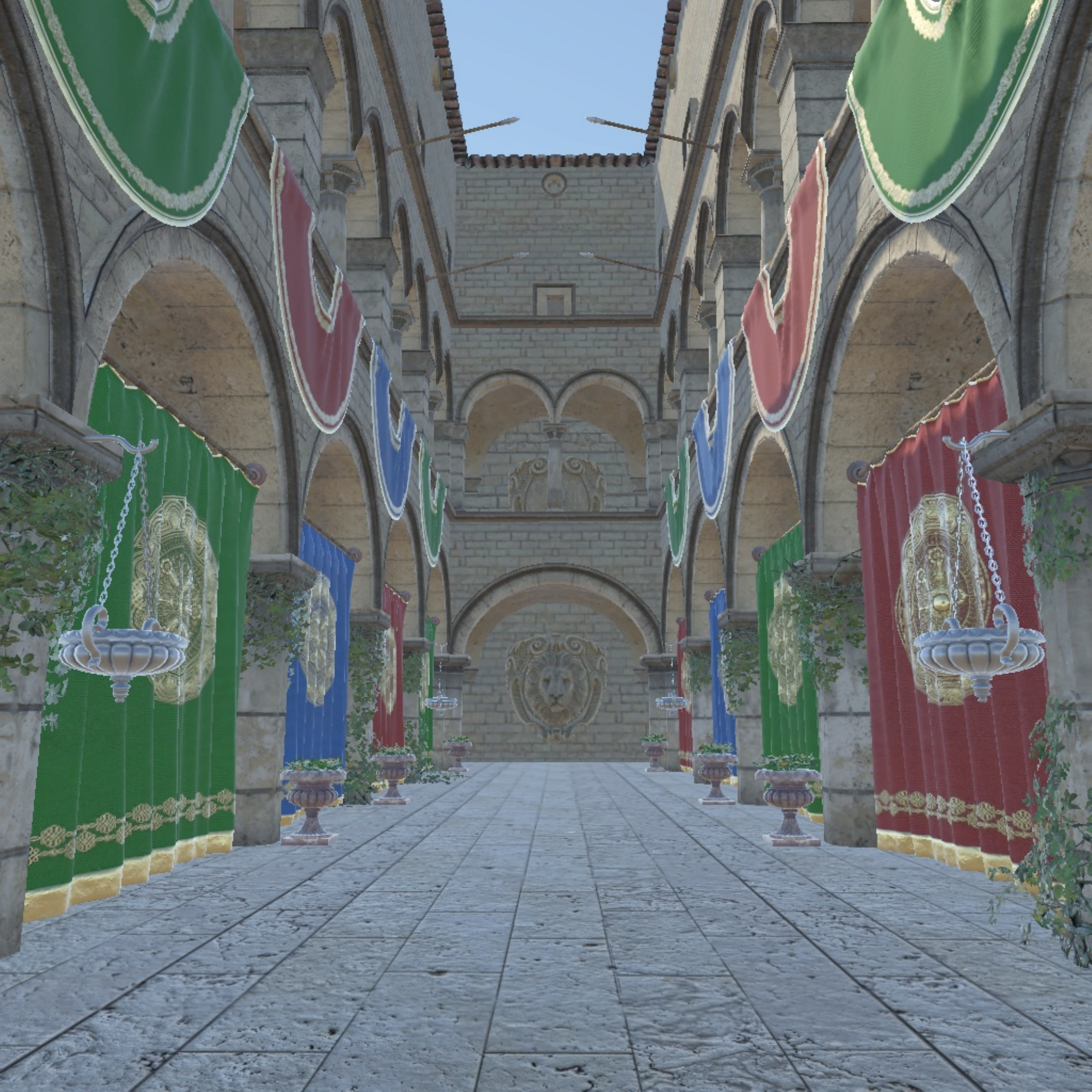}
    \caption*{Sponza (0.3M)}
  \end{subfigure}
  \caption{The five test scenes used in our evaluation, shown with their respective primitive counts.}
  \label{fig:scene_imgs}
\end{figure*}

We evaluate our method on two indoor scenes (Sponza, Robot Lab) and three outdoor scenes (Viking Village, Big City, Industrial Set v3.0), as illustrated in \Cref{fig:scene_imgs}. For each scene, we render a 60-second animation at 60 Hz along a pre-recorded camera path, resulting in 3600 frames per scene.

\subsection{Performance metrics}
\label{sec:method-metrics}

To evaluate the performance of the PVS estimation, we define the false negative rate FNR=FN/GTP ($\downarrow$), the false positive rate FPR=FP/GTP ($\downarrow$) and the pixel error rate PER ($\downarrow$) for comparison with previous work. The pixel error rate reports the pixels showing incorrect primitives because of primitives missed in the PVS. It is computed as a fraction of the screen resolution and averaged over a sequence of frames through the scene. Furthermore, we render shaded scenes from the PVS and the original geometry and compare the results using SSIM ($\uparrow$) \cite{wang_image_2004}, which considers local luminance, contrast, and structure over a sliding window.

\begin{figure*}[]
    \centering
    \includegraphics[width=\linewidth]{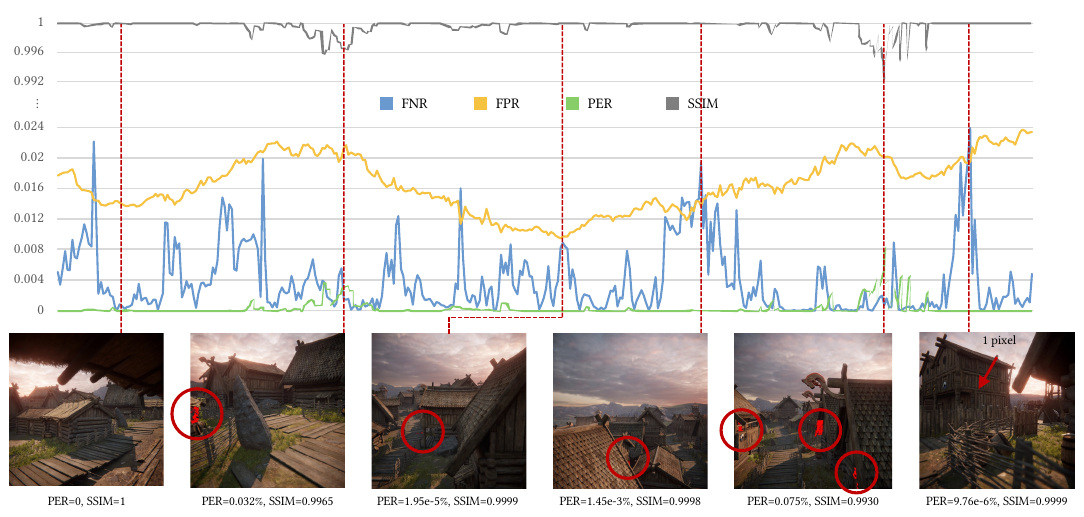}
    \vspace{-2.5em}
    \caption{Per-frame PVS estimation performance with key frames images of the Viking Village scene. The sequence has 1800 frames in total, with 410 frames of PVS computed shown in the figure. The error pixels of the key frames are marked in red on the rendered image.}
    \label{fig:per-frame-performance}
\end{figure*}

\Cref{fig:per-frame-performance} shows the metrics for an animated camera path of 3600 frames in the Viking village scene. For keyframes such as the best and the worst case of FNR, the rendering results of the computed PVS are shown to better illustrate the performance of the method. As indicated by the metrics and the rendered images, our method is robust in handling different geometry distributions and occlusion relations in the complex scene, with even the worst case in this challenging scene exhibiting only a small amount of pixel errors. The corresponding SSIM values confirm our observation that the error is usually unnoticeable to humans.


\Cref{fig:performance-by-r} shows the evaluation results on all scenes with different viewcell sizes $r$. We find that $r=30$ cm gives the best performance in all scenes, while $r=60$ cm has a slightly higher false negative rate. A potential explanation is that the size of the viewcell affects the subdivided grid shape in the geometry grid, where a too small grid size likely breaks the global geometry structure and causes the loss of global information, while a too large grid size makes it harder to learn the local structures within the grid.

\Cref{fig:performance-by-size} shows the evaluation results on all scenes with different interleaving grid size $d$. With different $d$ on the same scene, the performance can vary significantly. Choosing $d=16$ gives the best performance in most scenes. With a too large interleaving factor $d$, too few elements remain after downsampling in the CNN, which may cause the network to overfit and has a negative effect on performance. For smaller $d$, more elements remain after interleaving; assuming the same training period, the false negative will be higher due to greater difficulties of converging.

\Cref{tab:performance-by-d-ssim} shows the SSIM of all scenes with different $d$. SSIM is always very close to 1, which indicates that the rendering quality is very good. Additional perceptual (\scalebox{-1}[1]{F}LIP~\cite{andersson_flip_2020}) and temporal metrics (VMAF~\cite{li_toward_2016}, CGVQM~\cite{jindal_cgvqmd_2025}) are provided in the supplementary material, all confirming a similarly strong performance. Note that all our test scenes are completely unknown to the CNN, which is exclusively trained on purely synthetic scenes with simple, randomly generated geometries. Results show that our method is highly robust and generalizable to different and unknown data patterns while providing reliable estimation quality.


    \vspace{-2em}

\begin{figure*}[]
    \centering
    \includegraphics[width=0.94\linewidth]{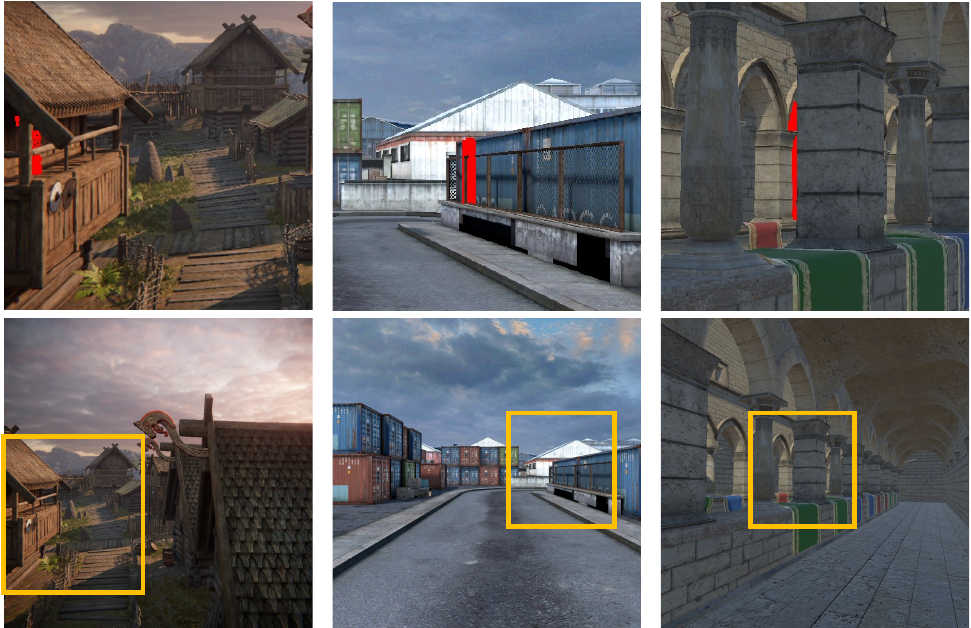}
    \vspace{-1em}
    \caption{Challenging situations for NeuralPVS model. left column: Multi-layer occlusion from Viking Village scene, 1104 false negative pixels. center column: complex occlusion structure from Industrial scene, 1152 false negative pixels. right column: almost occluded background from Sponza scene, 1102 false negative pixels.}
    \label{fig:challenging}
\end{figure*}


\subsection{Speed and memory}

We measure the runtime speed and memory in various scenes with different viewcell sizes. \Cref{fig:result-speed} presents the inference time per frame with the overall statistics and breakdown time per stage, including interleaving, CNN inference, and deinterleaving. The overall time per frame remains at approximately $10$ ms (100 Hz), despite the scene scale and geometry variations. This makes our method more efficient than a naive depth pre-pass, which is 1.5 ms slower in our experiments after both methods are averaged between frames.

Following the argument of Voglreiter et al.~\shortcite{voglreiter_trim_2023}, a running speed of 3 m/s translates into 5 cm/s camera movement at 60 Hz. At this speed of camera motion, the PVS is valid for 6 frames (100 ms),  when $r$=30 cm, and, for 18 frames (300 ms) when $r$=90 cm. Amortized over these valid periods, the PVS inference only takes approximately $3.3\%$ of the computation time for an application that generates new frames at a rate of 60 Hz. Therefore, we can assume that our method is very well suited for frame extrapolation or streaming applications in terms of speed.

\Cref{fig:performance-details-a} shows the time and peak memory allocated during inference. The adaptive relation convolution of OA-CNN will use more parameters if we add more feature channels, leading to increased inference time and memory usage when we use an interleaving factor of $d=32$. Although memory requirements can vary depending on the occupancy rate of the input $\mref{gv}$, the memory consumption is largely insensitive to scene geometry, as long as the viewcell configuration remains unchanged. The sparse tensor ensures that only the positions where geometry exists will be recorded in memory. Consequently, there is no memory overhead for empty grids.

\subsection{Comparisons to the state of the art}

To the best of our knowledge, the trim region method of Voglreiter et al.~\shortcite{voglreiter_trim_2023} is the fastest to date for from-region PVS computation. We compare our results with those reported in the original paper. For a fair comparison, we run an experiment using the same GPU (NVIDIA RTX 4090) and the same scenes as ours, kindly provided by the authors. \Cref{fig:comparison} shows the comparisons of metrics in these scenes, namely, Viking Village, Robot Lab, Sponza, and Big City. \Cref{tab:speed-comparison} shows the speed and memory comparisons on these scenes. 

Our method outperforms trim regions with respect to per-frame processing speed. The trim region method delivers $57$ Hz (18 ms) per frame on average, while we achieve over $100$ Hz (10 ms) on average (an improvement of $76\%$). We also obtain better quantitative performance, with significantly smaller false negatives and similar or better false positive rates. Compared to trim regions, our method has $83.8\%$ less FNR and $63.4\%$ less FPR on average. Although our PER is slightly higher, the overall image quality is not noticeably affected, as shown by the SSIM column in \Cref{tab:speed-comparison}.

\subsection{Ablation studies}

\label{sec:ablation-backbone}

To evaluate the effectiveness of the OA-CNN network and the interleaving function, we carried out experiments on the classical CNN backbone VNet \cite{milletari_v-net_2016} baseline and the original OA-CNN model with and without the interleaving function, as well as with and without RVL. Results are shown in \Cref{tab:ablation}. 

The model trained without RVL has a significantly higher false positive rate, which means that it simply predicts (almost) the whole scene as visible. The result shows that RVL is crucial in preventing the network from overfitting to the local minimum, and thus maintains a better balance between minimal FNR and reasonable FPR. Our results indicate that the interleaving function is the key to efficient inference. Its use increases the inference rate from 39 Hz to 100 Hz (2.5$\times$ speed-up) at 70\% reduced memory usage. 

The OA-CNN backbone is crucial for the performance of the network. Replacing OA-CNN with VNet leads to with 15\% increased FNR, while the FPR is almost unaffected. OA-CNN also contributes to decreasing inference time and memory usage. Without the interleaving function and OA-CNN backbone, a completely na\"ive VNet network is too slow for real-time applications.

\subsection{Challenging cases}

To better understand the limitations of our method, we list some typical challenging cases in \Cref{fig:challenging}. Like classical heuristic methods, the neural network also finds it harder to precisely estimate PVS in edge cases involving complex occlusion relationships (for example, multiple thin occluders in the left column). In the center column, the structure of the metal fence is not covered by our synthetic dataset and thus causes pixel errors. Further geometry has a lower resolution due to the geometry grid generation process, far-away geometry has a larger chance of getting misestimated, as shown in the right column. Moreover, machine learning methods are always dependent to some degree on the training data. While we found our randomized synthetic training set robust for a wide variety of scenes, fine-tuning our method for specific scenes can potentially help minimize errors.

\subsection{Dynamic scenes}

\begin{figure}
    \centering
    \includegraphics[width=\linewidth]{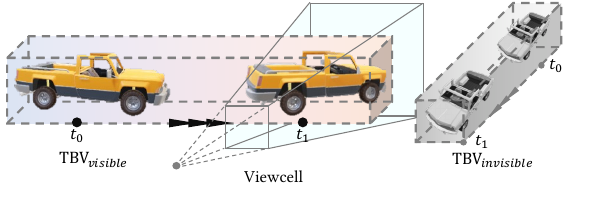}
    \caption{Dynamics scenes visibility computation with temporal bounding volumes. The yellow vehicle is visible during the period of time $t_0$ to $t_1$, while the grey vehicle is invisible.}
    \label{fig:dynamic-tbv}
    \vspace{-9pt}
\end{figure}

Since our view cell is defined purely in space, visibility changes caused by a dynamic object require special treatment. A simple approach unconditionally adds all dynamic objects to the PVS after visibility computation. However, this approach does not consider dynamic objects as occluders or occludees. While the first case---dynamic occluders---is likely too complicated to yield a speed-up, the second case---dynamic occludees---is rather simple to exploit. We construct a temporal bounding volume (TBV) for the moving object, which conservatively encloses the object during the predicted period (\Cref{fig:dynamic-tbv}). At runtime, the TBV is froxelized and tested against the froxelized PVS. If no froxel occupied by the TVB is deemed visible, the dynamic object can be pruned.

\section{Conclusion and future work}

We have presented a novel approach to compute from-regions potentially visible sets using a convolutional neural network. Our proposed method applies the OA-CNN network design with an additional interleaving function and a repulsive visibility loss to perform end-to-end estimation of the PVS. Our method achieves high accuracy and fast inference.

We see several directions for future work. With the development of deep learning, adopting a more featured network design would potentially bring better accuracy, while balancing the computational cost could be challenging. Moreover, additional spatial data structures, such as octrees or hash grids, might bring about further improvements in both performance and speed. We also consider replacing the simple froxelization with a more advanced feature representation of the local geometry. Finally, it is natural to expect some performance gain by introducing neural methods to other rendering tasks related to visibility, \eg radiance transfer, shadow rendering, and global illumination.

%


\input{main-arxiv.bbl}

\clearpage
\appendix
\setcounter{table}{0}
\renewcommand{\thetable}{S\arabic{table}}



\pagestyle{plain}                

\section*{Supplementary Material for \textit{NeuralPVS: Learned Estimation of Potentially Visible Sets}}



\renewcommand\footnotetextcopyrightpermission[1]{} 



\section{Implementation details}

\subsection{Voxelization}

Our voxelization procedure is described as follows:

\begin{lstlisting}[language=Python, basicstyle=\ttfamily\small]
for each voxel in volume parallel:
    worldPos = interpolate(volMin, volMax, voxelIdx, volDims)
    count = 0
    for each geometryPoint:
        if distance(worldPos, geometryPoint) < voxelSize:
            count += 1
    volume[voxelIdx] = count
\end{lstlisting}

Our framework supports both linear and logarithmic depth scaling. 
For simplicity, we adopt linear depth scaling in our implementation.

\subsection{Neural network}

Our implementation of the neural network is mainly based on the official implementation of OACNNs \cite{peng_oa-cnns_2024}, 
where we use the same parameters as theirs, while the last two layers of the CNN (out of four in total) are removed for faster inference. 
To be specific:

\begin{lstlisting}[language=Python, basicstyle=\ttfamily\small]
embed_channels = 64,
enc_num_ref = [16, 16],
enc_channels = [64, 64],
groups = [2, 4],
enc_depth = [2, 3],
down_ratio = [2, 2],
dec_channels = [96, 96],
point_grid_size = [[16, 32, 64], [8, 16, 24]],
dec_depth = [2, 2],
\end{lstlisting}

The number of channels depends on the interleaving grid size $d$. 
For example, if the interleaving grid size is $8$, the channel size will be $512$.

For the dense tensor version of the implementation, we use PyTorch. For the sparse tensor version, we use spconv \cite{spconv_contributors_spconv_2022}.

\section{Runtime breakdown}

As shown in \Cref{tab:time_breakdown}, the overhead introduced by NeuralPVS is minimal. The inference time matches the values reported in Table~3, while post-processing is effectively negligible since it consists only of a single index lookup in a custom lighting shader.

\begin{table}[h]
\centering
\caption{Runtime breakdown of PVS estimation time per frame.}
\label{tab:time_breakdown}
\begin{tabular}{@{}lc@{}}
\toprule
\textbf{Stage} & \textbf{Time (ms)} \\ \midrule
Pre-processing (voxelization) & 0.67 \\
Inference & 10.10 \\
Post-processing (pruning) & $<$ measurement limit \\
\textbf{Total} & \textbf{10.77} \\ \bottomrule
\end{tabular}
\end{table}

\section{Temporal consistency analysis}

To better evaluate the temporal consistency of a sequence rendered with our method, we adopt metrics including FLIP\cite{andersson_visualizing_2021}, VMAF~\cite{li_toward_2016} (temporal metric), and CGVQM~\cite{jindal_cgvqmd_2025} (temporal metric designed for full-sequence evaluation and sensitive to worst frames), as shown in \Cref{tab:more_metrics}.
The evaluation is based on a 60-second fresh render of $r=30, d=16$ with 60 FPS.


\begin{table}[H]  
  \centering
  \caption{Metrics for evaluating temporal consistency of the rendered sequences with NeuralPVS.}
  \label{tab:more_metrics}
  \resizebox{\columnwidth}{!}{%
    \begin{tabular}{@{}lccccccc@{}}
      \toprule
       & PSNR$\uparrow$ & SSIM$\uparrow$ & \scalebox{-1}[1]{F}LIP$\downarrow$ & VMAF$\uparrow$ & CGVQM$\uparrow$ & FNR$\downarrow$ & FPR$\downarrow$ \\ \midrule
      \textbf{Viking}     & 50.47 & 0.999 & 1.5e-04 & 99.99 & 99.92 & 0.0022 & 0.018 \\
      \textbf{Sponza}     & 92.23 & 1.000 & 3.9e-07 & 99.99 & 100.00 & 5.2e-4 & 0.026 \\
      \textbf{BigCity}    & 40.28 & 0.998 & 0.001   & 98.59 & 99.55 & 0.0023 & 0.018 \\
      \textbf{Robotlab}   & 40.44 & 0.998 & 0.001   & 97.85 & 99.54 & 0.0033 & 0.011 \\
      \textbf{Industrial} & 40.55 & 0.998 & 0.001   & 99.79 & 99.43 & 0.0040 & 0.005 \\ \bottomrule
    \end{tabular}
  }
\end{table}

\end{document}

%% file: utils/_def.tex
\usepackage{xstring}
\usepackage{xspace}
\usepackage[nameinlink]{cleveref}
\usepackage{subcaption}
\usepackage{xspace}
\usepackage{multirow}
\usepackage{graphicx}
\usepackage{float}       

\newcommand{\pvv}{PVS\xspace}

\makeatletter
\DeclareRobustCommand\onedot{\futurelet\@let@token\@onedot}
\def\@onedot{\ifx\@let@token.\else.\null\fi\xspace}

\def\eg{\emph{e.g}\onedot}

\makeatother

%% file: utils/math-macro.tex
\newcommand{\mref}[1]{%
    \IfEqCase{#1}{%
        {gv}{G(\mathbf{x})}
        {pvv}{V(\mathbf{x})}
        {gv_pt}{v^{g}}
        {pvv_pt}{v^{pv}}
        {pvv_pred}{\hat{V}}
        {pvv_pt_pred}{\hat{v^{pv}}}
        {pvv_prob}{\tilde{V^{pv}}}
        {pvv_pt_prob}{\tilde{v^{pv}}}
        {pvv_thres}{\tau}
        {idx_space}{\Omega}
        {neu_func}{f_\theta}
        {neu_interl}{g_r}
        {neu_deinterl}{g^{-1}_r}
        {neu_interl_r}{r}
        {neu_channel}{C}
        {gv_interl}{\hat{\mref{gv}}}
        {pvv_interl}{\hat{\mref{pvv_prob}}}
        {neu_func}{f_\theta}
        {coe_dice}{D_{dice}}
        {loss}{\mathcal{L}}
        {loss_dice}{\mathcal{L}_{dice}}
        {loss_dice_a}{\alpha_{dice}}
        {lossw}{\lambda}
        {loss_rv}{\mathcal{L}_{RV}}
        {loss_rv_attr}{\mathcal{L}_{RVAttr}}
        {loss_rv_attr_w}{\beta_{RV}}
        {loss_rv_rep}{\mathcal{L}_{RVRep}}
        {neu_tp}{\text{TP}}
        {neu_tn}{\text{TN}}
        {neu_fp}{\text{FP}}
        {neu_fn}{\text{FN}}
        {neu_fnr}{\text{FN}^*}
        {neu_fpr}{\text{FP}^*}
        {neu_fpratio}{\text{FPR}}
        {neu_fng}{\text{FNG}}
        {neu_ssim}{\text{SSIM}}
        {viewcell}{\Upsilon}
        {vc_origin}{\upsilon_o}
        {vc_left}{\upsilon_l}
        {vc_right}{\upsilon_r}
        {vc_pt}{\upsilon_*}
        {vc_radius}{r^\upsilon}
        {vc_fov}{\theta^\upsilon}
        {vc_z}{z^\upsilon}
        {vc_novel}{\hat\upsilon}
        {vc_plane}{p^\upsilon}
        {vc_frust}{F^\upsilon}
        {vc_frust_novel}{\hat{F^\upsilon}}
        {pvv_samp}{\mathcal V}
        {pvv_samp_num}{M}
        {pvv_samp_item}{\mathbf{v}'_m}
    }[\PackageError{mref}{Undefined option to math ref: #1}{}]%
}%